\documentclass[fleqn,12pt,epsfig]{article}
\usepackage{a4wide}
\usepackage{graphicx}
\newcommand{\df}[2]{\ensuremath{ {\raise
1pt\hbox{$\displaystyle #1$}\over \raise -2pt \hbox{$\displaystyle
#2$}}}}

\begin{document}
\def\d{{\rm d}}

\begin{titlepage}

\begin{center}
{\LARGE \bf On the existence of heavy tetraquarks%
\footnote{Based on a talk given at the
11th International Conference on Nuclear Reaction 
Mechanisms, Varenna (Italy), June 12-16, 2006}}\\[2cm]
{\large \bf{Fl.\ Stancu}\footnote{{\it e-mail}: fstancu@ulg.ac.be}}\\
{\small Institute of Physics, B.5, University of Liege,}\\
{\small Sart Tilman, B-4000 Liege 1, Belgium}\\%
\vspace{2cm}

{\normalsize \bf{In honor of David Brink} }\\

\end{center}
\vskip 1cm
\begin{abstract}\noindent

Previous work done in collaboration with David Brink 
is reviewed in the light of the recent observation of 
new charmonium-like resonances which can be interpreted as
tetraquarks.
In the framework of a schematic quark model
the spectrum of $c \bar c  q \bar q  $ tetraquarks is presented.
 \end{abstract}
\end{titlepage}

\section{Introduction}

The discovery by Belle \cite{Choi:2003ue}
of  the anomalously narrow $X(3872)$ meson
and of other hidden-charm states, in particular $X(3940)$ \cite{Abe:2004zs}
and $Y(4260)$ \cite{Aubert:2005rm}, revitalized the
experimental and theoretical interest in heavy quarkonium. 
The observation of $X(3872)$ has been confirmed by
CDF \cite{Acosta:2003zx}, DO \cite{Abazov:2004kp} and Babar 
\cite{Aubert:2004ns}. The difficulty of interpreting
these resonances as charmonium states are explained, 
for example, in Ref. \cite{QUIGG}. As alternatives,
these resonances might be interpreted as tetraquarks,
meson-meson molecules, hybrids, glueballs, etc. Here we shall
discuss the tetraquark option. 

The study of multiquark systems, as tetraquarks or hexaquarks 
has been initiated by Jaffe \cite{JAFFE} in the MIT bag model.
Later on it has been extended to potential models and applied to
light scalar mesons $a(980)$ and $f(980)$ \cite{WI83,Weinstein:1990gu}.
Subsequently heavy tetraquarks have been considered in search for
stability \cite{Zouzou:1986qh,Silvestre-Brac:1993ss}.
Flux tube models \cite{DOSCH,CP} suggested instability. 

In a previous work we studied a light tetraquark system of identical
particles in a standard nonrelativistic potential model
with colour confinement and hyperfine chromomagnetic interaction
\cite{Brink:1994ic}. Our result showed explicitly the role of 
hidden colour states (see below). 

Subsequently we have studied the stability of
a system of two light quarks $q = u, d$ and two heavy antiquarks $Q = c, b$ 
\cite{Brink:1998as}. Taking $Q = b$ we performed a simple variational
calculation with results comparable to other studies
\cite{Silvestre-Brac:1993ss}. We focused on the $qq \bar Q \bar Q$ 
system because it has more of a chance to be bound than 
$q \bar q Q \bar Q$, the latter having a lower threshold. The argument was 
based on the following inequality \cite{Nussinov:1999sx}
\begin{eqnarray}\label{ineg1}
m_{Q\overline{Q}}+ m_{q\bar{q}}&\le 2 m_{Q\bar{q}}~, 
\end{eqnarray}
valid for any value of the heavy-to-light mass ratio and where 
$m_{Q\bar{q}} = m_{q\bar{Q}} $. This means that 
$(Q\overline{Q})+(q\bar{q})$ is a lower threshold than 2 $(Q\bar{q})$. 
In addition  the $qq \bar Q \bar Q$ system is free of 
quark-antiquark annihilation processes 
which in  $q \bar q Q \bar Q$ cannot be avoided.

The open charm $cc \bar q \bar q$
tetraquark with spin $S=1$ and isospin $I=0$  appears unbound
in  Ref.~\cite{Silvestre-Brac:1993ss}, where the four-body problem was
solved by an expansion in a harmonic-oscillator basis up to $N=8$ quanta.
However, in the recent study of Janc and Rosina \cite{Janc:2004qn}
it is bound. 
This work, where a more sophisticated variational basis was used,
has been inspired by the considerations made 
in Ref. \ \cite{Brink:1998as} to include all possible channels
which accelerate the convergence and in particular the meson--meson 
channels.

In the above calculations  the potential of 
Bhaduri et al.\ \cite{Bhaduri:1981pn} has been used.
The hyperfine interaction is of chromomagnetic type.
The parameters include
constituent quark masses, the string tension of a
linear confinement, the strength of the Coulomb
interaction, and the strength and size parameter
of the hyperfine interaction which is a smeared
contact term. They were fitted over a wide range of 
mesons and baryons. 

In this paper we use the classification of tetraquark states 
as given in Refs. \ \cite{Brink:1994ic,Brink:1998as} and a simple
quark model to calculate the spectrum of $c \bar c q \bar q$ tetraquarks
and we discuss it in the light of recent data.

\section{The basis states}

Here we suppose that particles 1 and 2 are quarks and particles 3 and 4 
antiquarks, see Fig. 1.  Below we define the basis states 
in the orbital, colour and spin space taking into account the Pauli principle.
The flavour space is trivial if we restrict to SU(2).
The total wave function of a tetraquark is a linear combination of products of
orbital, spin, flavour and colour parts.

\subsection{The orbital part} 
There are at least three possible ways to define the relative
coordinates. The three relevant possibilities for our problem are
shown in Fig. 1. 
In the cases $(a),(b)$ and $(c)$ the internal coordinates are 
\begin{equation}
   \vec{\sigma}  = \frac{1}{\sqrt{2}} (\vec{r_1}-\vec{r_2}),
~~~\vec{\sigma'} = \frac{1}{\sqrt{2}} (\vec{r_3}-\vec{r_4}),
~~~\vec{\lambda} = \frac{1}{2}(\vec{r_1}+\vec{r_2}-\vec{r_3}-\vec{r_4}),   
\end{equation}
\begin{equation}\label{direct}
   \vec{\rho}  = \frac{1}{\sqrt{2}} (\vec{r_1}-\vec{r_3}), 
~~~\vec{\rho'} = \frac{1}{\sqrt{2}} (\vec{r_2}-\vec{r_4}),
~~~\vec{x}     = \frac{1}{2}(\vec{r_1}-\vec{r_2}+\vec{r_3}-\vec{r_4}),   
\end{equation}
\begin{equation}\label{exchange}
   \vec{\alpha}  = \frac{1}{\sqrt{2}} (\vec{r_1}-\vec{r_4}), 
~~~\vec{\alpha'} = \frac{1}{\sqrt{2}} (\vec{r_2}-\vec{r_3}),
~~~\vec{x}     = \frac{1}{2}(\vec{r_1}-\vec{r_2}-\vec{r_3}+\vec{r_4}).   
\end{equation}
The first system of coordinates is convenient when the quarks or 
antiquarks are correlated to form diquarks, as  
in the diquark-antidiquark model. The coordinates (2) and (3) 
are useful in describing the direct and exchange meson-meson 
channels \emph{ i. e.} the strong decays. One should
use the system which is more appropriate for a given problem. But in 
specific calculations one can pass from one coordinate system to the other
by orthogonal transformations.
In the diquark-antidiquark picture
the radial wave function can be written as a function of 6 variables
$R({\sigma}^2,  {\sigma'}^2, {\lambda}^2, \vec{\sigma} \cdot \vec{\sigma'},
\vec{\sigma} \cdot \vec{\lambda}, \vec{\sigma'} \cdot \vec{\lambda})$
but it can also be expressed in terms of coordinates (\ref{direct})
or (\ref{exchange}).
All channels which accelerate convergence should be included. The 
non-orthogonality between different $R$ functions can be handled without 
problems\ \cite{Brink:1994ic,Brink:1998as}.

\begin{figure}\label{fig2}
\begin{center}
\includegraphics*[width=10.0cm,keepaspectratio]{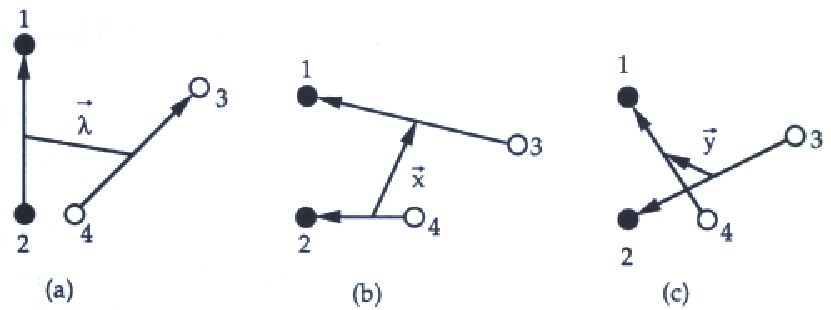}
\end{center} 
\caption{Three independent relative coordinate systems. Solid and 
open circles represent quarks and antiquarks respectively: (a)
diquark-antidiquark channel, (b) direct meson-meson channel, (c) exchange
meson-meson channel. }
\end{figure}

\subsection{The colour part}
In the colour space one can construct a colour singlet tetraquark 
state using intermediate couplings associated to the three
coordinate systems defined above. In this way one obtains three 
distinct bases. These are
\begin{equation}\label{diquark}
|\overline{3}_{12} 3_{34} \rangle, ~~~ |6_{12} \overline{6}_{34} \rangle,
\end{equation}
\begin{equation}\label{directcolor}
|1_{13} 1_{24} \rangle, ~~~ |8_{13} 8_{24} \rangle,
\end{equation} 
\begin{equation}\label{exchangecolor}
|1_{14} 1_{23} \rangle, ~~~ |8_{14} 8_{23} \rangle.
\end{equation} 
The 3 and $\overline{3}$ are antisymmetric and 6 and $\overline{6}$
are symmetric under interchange of quarks or antiquarks.
Each set,  (\ref{directcolor}) or  (\ref{exchangecolor}), 
contains a singlet-singlet colour
and an octet-octet colour state. The amplitude of the latter
vanishes asymptotically, when the mesons, into which a tetraquark 
decays, separate. These
are called \emph{hidden colour} states by analogy to states
which appear in  the nucleon-nucleon problem, defined as a six-quark system
\ \cite{HARVEY}.
The contribution of hidden colour states to the 
binding energy of light tetraquarks has been calculated explicitly in 
Ref.  \cite{Brink:1994ic}.
Below we shall point out their crucial role in the description 
of $c \bar c q \bar q$ tetraquarks.

\subsection{The spin part} 

As the quarks and antiquarks are spin 1/2 particles the total 
spin of a tetraquark can be $S = 0, 1$ or 2. 
For $S = 0$ there are two independent basis states for each channel.
The bases associated to 
(\ref{diquark}), (\ref{directcolor}) and  (\ref{exchangecolor}) are
\begin{equation}\label{diquark0}
|\chi_+ \rangle, ~~~ |\chi_- \rangle,  
\end{equation}
\begin{equation}\label{directcolor0}
|P_{13} P_{24} \rangle,~~~ |(V_{13} V_{24})_0 \rangle, 
\end{equation} 
\begin{equation}\label{exchangecolor0}
|P_{14} P_{23} \rangle,~~~ |(V_{14} V_{23})_0 \rangle, 
\end{equation} 
where $P$ and $V$ stand for pseudoscalar and vector meson 
subsystems respectively and the lower index 0 indicates the total spin.
The corresponding Young tableaux for the states (\ref{diquark0}) are
shown in Fig. 2. 
\begin{figure}\label{fig2}
\begin{center}
\includegraphics*[width=6.0cm,keepaspectratio]{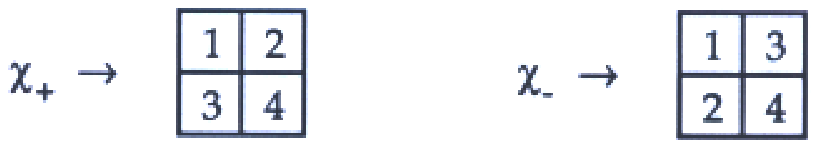}
\end{center} 
\caption{The Young tableaux corresponding to the states (\ref{diquark0}). }
\end{figure}

For $S = 1$ there are three independent states
in each channel, to be identified by three
distinct Young tableaux \cite{book}. 
As an example we give the basis 
for the direct meson-meson channel  \cite{Brink:1998as}
\begin{equation}\label{directcolor1}
|(P_{13} V_{24})_1 \rangle, ~~~~|(V_{13} P_{24})_1 \rangle, 
~~~|(V_{13} V_{24})_1 \rangle.
\end{equation}
The lower index indicates the total spin 1.
The permutation property under transposition (13)  manifestly is
\begin{equation}
    (13) |P_{13} \rangle = - |P_{13} \rangle,
~~~ (13) |V_{13} \rangle = + |V_{13} \rangle,
\end{equation}
and similarly for (24)
\begin{equation}
    (24) |P_{24} \rangle = - |P_{24} \rangle,
~~~ (24) |V_{24} \rangle = + |V_{24} \rangle.
\end{equation}
With the identification 1 = $c$, 2 = $q$, 3 = $\overline c$ and 
4 = $\overline q$ the permutation (13)(24) is equivalent to 
charge conjugation, reason for which the basis vectors
$|\alpha_3 \rangle$ and $| \alpha_6 \rangle $ defined below
have charge conjugation $C = 1$ and $|\alpha_1 \rangle$, $|\alpha_2 \rangle$,
$|\alpha_4 \rangle$, $|\alpha_5 \rangle$ have charge conjugation $C = - 1$.
  
The case $S = 2$ is trivial. There is a single basis state 

\begin{equation}\label{SPIN2}
\chi^S = |(V_{13} V_{24})_2 \rangle , 
\end{equation}
which is symmetric under any permutation of quarks.

\section{Models for X(3872)}

Below we shall first briefly describe the diquark-antidiquark 
model and  next we shall calculate the full spectrum of
tetraquarks $c \overline c$$q \overline q$ in the 
framework of a simple model.

\subsection{The diquark-antidiquark model}

The diquark model for tetraquarks, named also exotic mesons,
has been attractive for some time already. One obstacle is to estimate 
the mass of the diquark. A simple consistent picture has been given
by Lichtenberg et al. \cite{Lichtenberg:1996fi}. In that context the mass of 
$q q \bar c \bar c$ with diquarks of spin 0 and 1 was 3910 MeV
and that of $q c \bar q \bar c$ with diquarks of spin zero 3920 MeV.

The heavy-light diquark picture of Maiani et al. 
\cite{Maiani:2004vq} is more sophisticated. It contains explicitly
a hyperfine interaction between quarks (antiquarks) and the diquarks 
have spin zero or one in all cases. However the colour space is truncated
to $\overline 3 3$ 
so that the number of basis vectors is twice less than that  
presented in the following subsection
\footnote{The truncation of the colour space is questionable 
for tetraquarks involving charmed quarks in view of the 
results of Refs. \cite{Zouzou:1986qh}. For tetraquarks involving $b$ quarks
the approximation is better.  }.  
Accordingly there are only two basis vectors for
$J^{PC} = 0^{++}$, one for  $J^{PC} = 1^{++}$, two for  $J^{PC} = 1^{+-}$
and one for $J^{PC} = 2^{++}$, so that in the tetraquark spectrum there are 
twice less states than in Fig. 3 below. The narrow decay width is 
explained by a rearrangement and by a "bold guess" of the 
coupling constant $g_{X J/\psi V}$. 

\subsection{The role of octet-octet channels}

Here we  use the simple model of Ref. \cite{Hogaasen:2005jv}
in order to point out the important role of hidden colour channels
introduced above. 
Accordingly the mass of a tetraquark is given by 
\begin{equation}\label{eq:Mcm}
\mathcal{M} =\sum_i m_i + \langle H_{\mathrm{CM}} \rangle,
\end{equation}
where

\begin{equation}\label{eq:Hcm}
H_{\mathrm{CM}} = - \sum_{i,j} C_{ij}\,
~\lambda^c_{i} \cdot \lambda^c_{j}\,\vec{\sigma}_i \cdot 
\vec{\sigma_j}~.
\end{equation}
The first term in Eq. (\ref{eq:Mcm}) contains the effective masses $m_i$ which 
incorporate the conventional constituent mass plus the kinetic 
energy and the confinement potential contributions. The constants $ C_{ij}$
represent the integral in the coordinate space of some unspecified 
analytic form of the one gluon-exchange potential and of spatial 
wave functions.
A fit within $\pm10$ MeV to properties of charmed baryons gave the
following parameters 

\begin{equation}\label{eq:par}
\begin{array}{lll}
  C_{qq}=20\,\mathrm{MeV},&
C_{qc}=5\,\mathrm{MeV},&C_{qs}=15\,\mathrm{MeV},\\
C_{ss}=10\,\mathrm{MeV},&
C_{cs}=4\,\mathrm{MeV}, &C_{c\overline{c}}=4\,\mathrm{MeV}.
\end{array}\end{equation}

For a tetraquark the possible states are
$J^{PC} = 0^{++}, 1^{++}, 1^{+-}$ and $J^{PC} = 2^{++}$.
In each case a basis can be built
with (1,3) and (2,4) as quark-antiquark subsystems where each
subsystem has a well defined colour,  singlet or octet. 
This arrangement corresponds to $J/\psi $ + \emph{light meson} channel.
Other intermediate couplings can also be defined, as
for example, (1,4) and (2,3) which can correspond to the
$D + {\overline D}^*$ channel, when the total spin allows. 
One can pass from one 
coupling to the other, depending on  
the problem one looks at and also, for convenience in the calculations. 

For the colour-spin basis states we shall use the 
notation introduced in the previous section, i. e. $1_{mn}$ 
for colour singlet and $8_{mn}$ 
for colour octet subsystems and 
$P_{mn}$ and $V_{mn}$ for spin 0 and 1 respectively.

For $J^{PC} = 0^{++}$ the basis constructed from products 
of states (\ref{directcolor}) and (\ref{directcolor0}) is
\begin{eqnarray}\label{eq:betai}
\hskip -10pt& \gamma_1 = 
| 1_{13} 1_{24} P_{13} P_{24} \rangle,\
&  \gamma_2 =
| 1_{13} 1_{24} (V_{13} V_{24})_0 \rangle, \nonumber\\
\hskip -10pt& \gamma_3 =
| 8_{13} 8_{24} P_{13} P_{24} \rangle,\ 
&\gamma_4 =
| 8_{13} 8_{24}  (V_{13} V_{24})_0 \rangle .
%
%
\end{eqnarray}
The chromomagnetic interaction Hamiltonian with minus sign, -$H_{\mathrm{CM}}$,
acting on this basis leads to the following symmetric matrix
\\ 
\vskip -20pt
$$\left[
\renewcommand{\arraystretch}{2.0}
\begin {tabular}{cccccc}
%
$16(C_{13}+C_{24})$&0&0&$ 8 \sqrt{\df{2}{3}}(C_{12}+C_{23})$\\
%
 &$-
\df{16}{3}(C_{13}+C_{24})$&$ -8 \sqrt{\df{2}{3}}(C_{12}+C_{23})$
&-$\df{16 \sqrt2}{3}(C_{12}-C_{23})$\\
%
%
 & &$-2(C_{13}+C_{24})$ & $\df{4}{\sqrt3}(2C_{12}-7C_{23})$\\
%
%
 & & & $\df{16}{3}C_{12}+\df{56}{3}C_{23}+\df{2}{3}(C_{13}+C_{24})$
\\
\end{tabular}\right]$$
%
For $J^P=1^+$ there are six linearly independent basis 
vectors built as products of colour (\ref{directcolor}) and 
spin (\ref{directcolor1}) states.
\begin{eqnarray}\label{eq:alphai}
\hskip -10pt& \alpha_1 =
| 1_{13} 1_{24} (P_{13} V_{24})_1 \rangle,\ 
&  \alpha_2 =
%
| 1_{13} 1_{24} (V_{13} P_{24})_1 \rangle,\nonumber\\
\hskip -10pt& \alpha_3 =
| 1_{13} 1_{24}~ (V_{13} V_{24})_1 \rangle, \  
&\alpha_4 =
%
| 8_{13} 8_{24} (P_{13} V_{24})_1 \rangle, \nonumber\\
\hskip -10pt& \alpha_5 =
| 8_{13} 8_{24} (V_{13} P_{24})_1 \rangle, \ 
& \alpha_6 =
| 8_{13} 8_{24}~ (V_{13} V_{24})_1 \rangle.
\end{eqnarray}
from which $\alpha_3$ and $\alpha_6$ have charge conjugation
$C = 1$ and the others  $C = - 1$. The 6 $\times$ 6
matrix of -$H_{\mathrm{CM}} $
can be found in Ref. \cite{Hogaasen:2005jv}. Note that all matrix elements 
between states with opposite charge conjugation are naturally zero which
means that the 6 $\times$ 6 matrix of Ref. \cite{Hogaasen:2005jv}
has a block-diagonal form. One block is 
the 2 $\times$ 2 submatrix for states of charge conjugation $C = 1$.
This is 
\\
\vskip -20pt
$$\left[
\renewcommand{\arraystretch}{2.0}
\begin {tabular}{cc}
%
$ -\df{16}{3}(C_{13}+C_{24})$ & $ -{\df{8 \sqrt2}{3}}(C_{12}-C_{23})$\\
%
 & $ \df{2}{3} (4C_{12}+14C_{23}+C_{13}+C_{24})$ \\
\end{tabular} \right]$$
which can be related to X(3872). The other
block is a 4 $\times$ 4 submatrix  for states of charge conjugation $C = - 1$, 
not written explicitly here, but which can be easily identified
from Ref.  \cite{Hogaasen:2005jv}.

For $J^{PC} = 2^{++}$ the basis vectors are
\begin{eqnarray}\label{eq:deltai}
\hskip -10pt& \delta_1 =
| 1_{13} 1_{24} \chi^S \rangle, \ 
&  \delta_2 =
| 8_{13} 8_{24} \chi^S \rangle, 
\end{eqnarray}
where $\chi^S$ is the $S = 2$ spin state (\ref{SPIN2}).
The corresponding 2 $\times$ 2 matrix is
\\
\vskip -20pt
$$\left[
\renewcommand{\arraystretch}{2.0}
\begin {tabular}{cc}
%
$ -\df{16}{3}(C_{13}+C_{24})$ & $ {\df{8 \sqrt2}{3}}(C_{12}-C_{23})$\\
%
 & $ - \df{2}{3} (4C_{12}+14C_{23}-C_{13}-C_{24})$ \\
\end{tabular} \right]$$
In the calculation of the matrix elements we have used the equalities
\begin{equation}
C_{14} = C_{23}, ~~~C_{12} = C_{34},
\end{equation}
due to charge conjugation.

One can extend the observation made in Ref. \cite{Hogaasen:2005jv},
for $1^{++}$  to $2^{++}$ states as well, namely
the matrices of both $1^{++}$ and $2^{++}$ are diagonal 
provided the chromomagnetic interaction is the same
for a quark-quark pair as for a quark-antiquark pair, ${\it i. e.}$
$C_{12}$ = $C_{23}$.
This implies that the eigenvectors are either a pure colour singlet-singlet or 
a pure colour octet-octet state. In each case the pure  octet-octet state is 
the lowest  and obviously cannot dissociate into a charmonium state and a 
vector meson.
But if one allows for a difference between  $C_{12}$ and $C_{23}$, see
parameters (\ref{eq:par}), then the lowest state receives
a small singlet-singlet component which can then decay into
$J/\psi + \rho$ or  $J/\psi + \omega$ with a small width. 
Such a description is very suitable for $X(3872)$ which has a width 
$\Gamma  < 2.3$ MeV (95 \% C.L.) \cite{Choi:2003ue}.

The states with $J^P = 1^{+-}$ can decay into  $J/\psi$ + 
\emph{pseudoscalar meson}
as required by charge conjugation.
There is no observation which can be associated to these states.

In Fig. 3 we show the calculated spectrum with the 
parameters (\ref{eq:par}). These parameters were obtained from a
fit to charmed baryons, but determined within $\pm$10 MeV. This means that 
the levels in Fig. 3 should be submitted to some uncertainties,
not easy to be estimated.
One can see that the 3910 MeV $1^{++}$ level is close to 
$X(3872)$  but the 4057 MeV $2^{++}$ level is about 100 MeV above the observed
$X(3940)$ resonance. 
\begin{figure}\label{fig4}
\begin{center}
\includegraphics*[width=10.0cm,keepaspectratio]{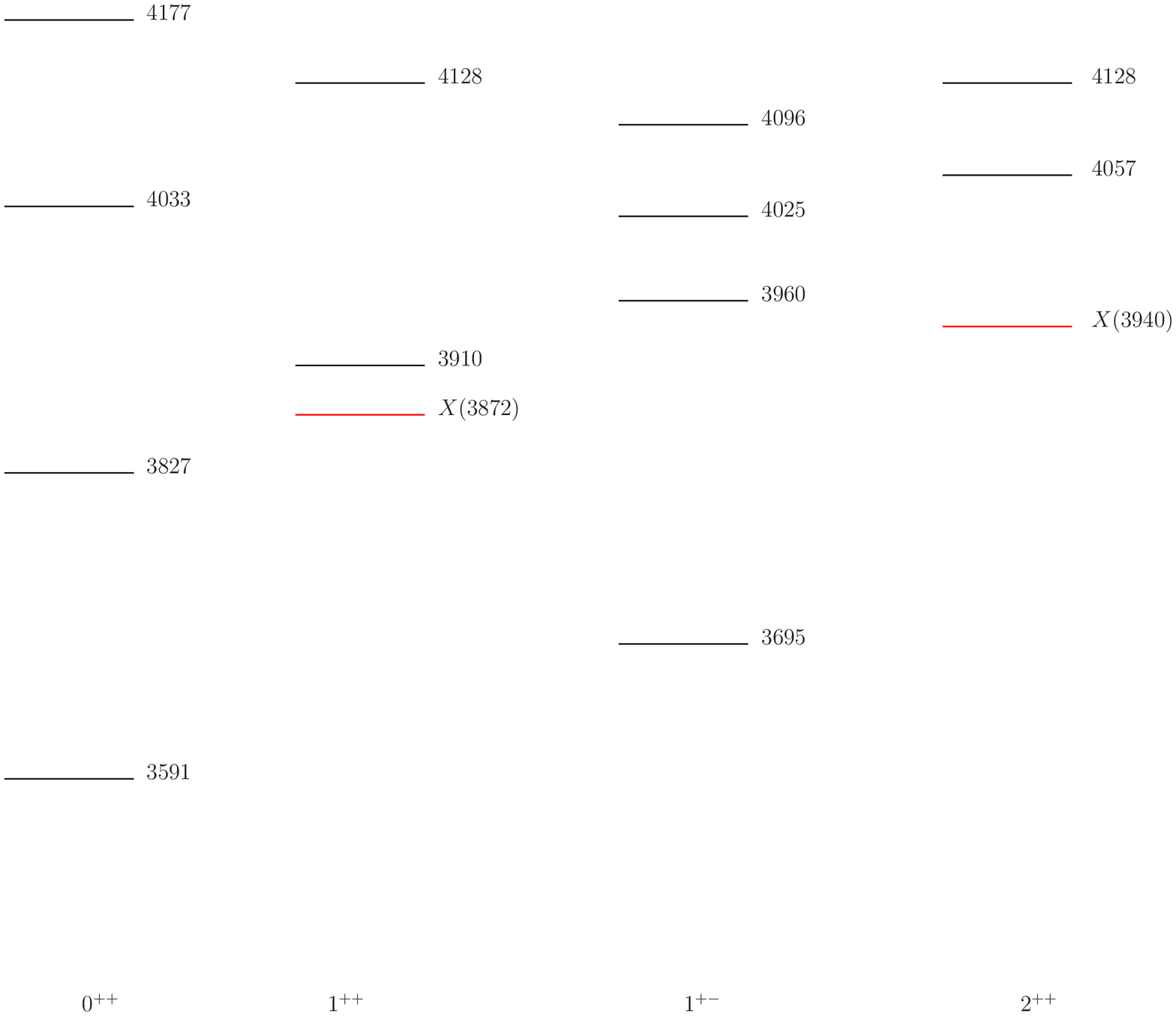}
\end{center} 
\caption{The spectrum of $c \overline c q \overline q$ tetraquarks}
\end{figure}

\section{Conclusions}\label{se:op}

The work of Ref. \cite{Hogaasen:2005jv} and its present extension 
suggests that the resonance X(3872) is an interesting candidate for
the tetraquark $c \overline c q \overline q$. 
But the resonance X(3940) does not fit so well into this scheme
as a $J^P = 2^{++}$ state, its mass being 100 MeV below  
the calculated value. 
For X(3872) further work is necessary. One has to quantitatively estimate
the contribution of annihilation channels. Also estimates  
of the strong decay widths are necessary before any definite
conclusion.

Besides the chromomagnetic interaction, other effects may be important to 
the description
of tetraquarks, as for example a long-range strong attraction
between light quarks, generated by one-meson exchange. Actually this is
the basic mechanism of the molecular picture \cite{TORNQVIST,Swanson:2003tb}.
A speculative discussion on an additional effect of the one-meson
exchange interaction can be found in Ref.
\cite{Richard:2005jz}


\end{document}